\documentclass[aps,pre,showpacs,showkeys,twocolumn,preprintnumbers,floatfix,nofootinbib,10pt]{revtex4-2}
\usepackage{amssymb}
\usepackage{graphicx,color}
\usepackage{multirow}

\begin{document}

\title{Unveiling the hidden weak universality of the ZGB model}
\author{Henrique A. Fernandes$^{1}$, Roberto da Silva$^{2}$}

\affiliation{$^1$Instituto de Ci{\^e}ncias Exatas e Tecnol{\'o}gicas, Universidade Federal de Jata{\'i}, BR 364, km 192, 3800 - CEP 75801-615, Jata{\'i}, Goi{\'a}s, Brazil \\
$^2$Instituto de F{\'i}sica, Universidade Federal do Rio Grande do Sul, Av. Bento Gon{\c{c}}alves, 9500 - CEP 91501-970, Porto Alegre, Rio Grande do Sul, Brazil}

\begin{abstract}

In this work, we revisited the Ziff-Gullari-Barshad (ZGB) model in order to investigate its critical behavior when carbon monoxide (CO) molecules are allowed to desorb from the catalytic surface. As shown by several authors, when this kind of desorption takes place, the first-order phase transition of the standard model disappears, and an Ising-like critical point is found for a very small value of the desorption rate. However, our time-dependent Monte Carlo simulations reveal that, instead of a single critical point, there exists a critical line that encompasses multiple universality classes, passing through the three- and four-state Potts points, as well as, the Ising one, resulting in an unprecedented critical line of weak universality.

\end{abstract}

\maketitle

\section{Introduction}

\label{sec:introduction}

The ZGB model, originally devised by R. M. Ziff, E. Gulari, and Y. Barshad
in 1986 \cite{ziff1986}, is one of the most extensively studied catalytic
surface models, primarily due to its intriguing property of exhibiting both
continuous and discontinuous phase transitions. The model describes the
production of carbon dioxide (CO$_{2}$) molecules through the reaction of
carbon monoxide (CO) and oxygen atoms (O) on a catalytic surface. This
dual-phase transition behavior makes the ZGB model a fascinating subject of
study in surface reaction dynamics.

It is worth highlighting the remarkably short time in which the model gained
significance in its field. Since then, researchers worldwide, employing various techniques, and introducing numerous modifications to the original work. These adaptations are thoroughly documented in the literature, including studies such as \cite{meakin1987,dickman1986,fischer1989,marro1999, tome1993,dumont1990,albano1992,kaukonen1989,jensen1990,brosilow1992,matsushima1979, buendia2013,buendia2015,chan2015,ehsasi1989,grandi2002,hoenicke2000,buendia2012, hoenicke2014,satulovsky1992,albano1990,brosilow1993}.

In this paper, we report our findings about the ZGB model with CO desorption
rate $k$, considering a specific region of its phase diagram close to the
point of the discontinuous phase transition of the standard model for small
values of $k$. As we show, our results support the existence of an
unprecedented critical line in the $y\times k$ space, where $y$ is the CO-adsorption rate, which extends from $0.525\lesssim y\lesssim 0.560$ and $0.01\lesssim k\lesssim 0.08$. By analyzing some points, we will be able to conclude that the model possesses multiple universality, since the dynamic critical
exponent $\theta $ varies along this critical line. Furthermore, we will
show that this critical line comprises points belonging to the different
universality class of very important models such as the Ising model and the three-
and four-state Potts models.

As we have presented in Figs. 2 and 4, as well as in Table 1 of Ref. \cite{fernandes2018}, the continuous phase transition of the standard ZGB model
remains unchanged for increasing CO desorption rate values $k$, showing that the transition from the O-poisoned state to the active phase is not affected by the desorption of CO molecules. However, the
discontinuous phase transition of the original version of the model, which
takes place at $y \simeq 0.525$ and separates the active phase from the
CO-poisoned state, is destroyed. An easy explanation for this fact is that
the desorption of CO molecules ($k > 0$) makes the existence of
this poisoned state impossible. Here, a poisoned state means an absorbing phase, which occurs when all sites of the lattice are filled with only O atoms or CO
molecules.

In this way, the ZGB model with CO desorption becomes part of a select group
of models that possess weak universality, such as the Ashkin-Teller model 
\cite{ashkin1943}, Baxter model \cite{baxter1971}, as well as those
presented in Refs. \cite{fabien2005, malakis2009, queiroz2011, jin2012,
sau2023,Silva2014,Fernandes2017}.

In the next section, we present the ZGB model modified to allow the
desorption of CO molecules as well as the details of the nonequilibrium
Monte Carlo method considered in our simulations, and the refinement based
on the coefficient of determination technique, used for the first time to
find critical points of the generalized Ising model \cite{silva2012}, but
that had fruitful results in many other contexts, as in determining some
important critical points as the Lifshitz one in the ANNNI model \cite{silva2013a} and the tricritical point in the metamagnetic model \cite{silva2013b}. Section \ref{sec:results} is devoted to the presentation of the results
obtained in this study. In that section, we show the critical line present
in a specific region of the phase diagram of the model, along with some
critical exponents which confirm the weak universality of the modified
model. In Sec. \ref{sec:conclusions}, we summarize and conclude our work.

\section{Model and simulation method}

\label{sec:model}

ZGB model \cite{ziff1986} mimics, in a catalytic surface, the adsorption of
carbon monoxide (CO) and oxygen ($\text{O}_{2}$) molecules along with the
desorption of carbon dioxide ($\text{CO}_{2}$) molecules from that surface
soon after the reaction of a CO molecule with a neighboring O atom.

In this model, the catalytic surface is represented by a two-dimensional
regular square lattice by following a Langmuir-Hinshelwood mechanism \cite{ziff1986, evans1991b}. These processes can be represented by the following
reaction equations: 
\begin{equation}
\text{CO}(g)+V\longrightarrow \text{CO}(a)  \label{eq:co_ad}
\end{equation}%
\begin{equation}
\text{O}_{2}(g)+2V\longrightarrow 2\text{O}(a)  \label{eq:o2_ad}
\end{equation}%
\begin{equation}
\text{CO}(a)+\text{O}(a)\longrightarrow \text{CO}_{2}(g)+2V
\label{eq:co2_des}
\end{equation}%
where $g$ refers to molecules in the gas phase, $a$ refers to
molecules/atoms adsorbed on the surface, and $V$ stands for vacant sites.

As can be seen, Eqs. (\ref{eq:co_ad}) and (\ref{eq:o2_ad}) are related to
the adsorption processes. In the first one, a CO molecule is selected in the
gas phase with a rate $y$ and a site on the surface is chosen at random. If
that site is vacant $V$, the molecule is immediately adsorbed on it.
Otherwise, if the chosen site is occupied, the CO molecule returns to the
gas phase and the trial ends. In the second one, an $\text{O}_{2}$ molecule
in the gas phase strikes the surface with a rate $1-y$ and a
nearest-neighbor pair of sites is chosen at random. If both sites are vacant
($2V$), the $\text{O}_{2}$ molecule dissociates into a pair of $O$ atoms
which are adsorbed on those sites. On the other hand, if one or both sites
are occupied, the $\text{O}_{2}$ molecule returns to the gas phase and the
trial ends. The desorption process occurs according to Eq. (\ref{eq:co2_des}) and is considered after each successful adsorption process. In that case,
all nearest-neighbor sites of a newly occupied site are checked randomly
and, if one $\text{O-CO}$ pair is found, they react immediately forming a $\text{CO}_{2}$ molecule which desorbs leaving behind two vacant sites.

As depicted above, the model possesses only one control parameter, the
CO-adsorption rate $y$, which is used to determine its two irreversible
phase transitions, one continuous and another discontinuous, separating
three distinct states: one O-poisoned state that remains for $0\leq y<y_{1}$, one CO-poisoned state which occurs when $y_{2}<y\geq 1$, and one active
phase with sustainable production of $\text{CO}_{2}$ molecules for $y_{1}<y<y_{2}$. The transition points, $y_{1}$ and $y_{2}$, are well-known in literature with the continuous phase transition occurring when $y=y_{1}\approx 0.3874$ \cite{voigt1997} and the discontinuous one for $y=y_{2}\approx 0.5256$ \cite{ziff1992}.

The ZGB model studied in this work is modified to allow the spontaneous
desorption of CO molecules from the catalytic surface with a rate $k$ \cite{tome1993,fischer1989,dumont1990, albano1992, kaukonen1989, jensen1990,
brosilow1992, matsushima1979, buendia2013, buendia2015,
chan2015,fernandes2018}. The equation addressing this process is given by 
\begin{equation}
\text{CO}(a)\longrightarrow \text{CO}(g)+V,  \label{eq:co_des}
\end{equation}
i.e., a CO molecule adsorbed on the surface returns to the gas phase and
leaves behind an empty site.

The inclusion of this equation to those described above takes into account
the results obtained in experiments in which the desorption of CO molecules
from the catalytic surface occurs without any reacting process. On the other
hand, we do not consider the spontaneous desorption of O atoms as the rate
in which this process occurs is much smaller than for the CO molecule \cite{ehsasi1989}.

As presented in Ref. \cite{fernandes2018}, in this work we consider the density of vacant sites $\rho $ as the order parameter of the model. It is defined as 
\begin{equation}
\rho (t)=\frac{1}{L^{2}}\sum_{i=1}^{L^{2}}s_{i},  \label{eq:density}
\end{equation}%
where $t$ is the Monte Carlo (MC) step of our nonequilibrium simulations, $L$
is the linear size of a regular square lattice, and $s_{i}$ is equal to 1
when the site $i$ is vacant, otherwise, it is equal to zero.

In this study, the results were obtained using the well-established
short-time (nonequilibrium) Monte Carlo (MC) method. For spin systems,
time-dependent MC simulations are grounded in the seminal work of Janssen,
Schaub, and Schmittmann \cite{janssen1989}, which applied renormalization
group theory within field theory. This was later supported numerically by
Huse \cite{huse1989}. Their work demonstrated that universality and scaling
behavior emerge even in the early stages of time evolution in these systems,
with considerations of the randomness in the evolution and the dependence on
initial conditions. Since then, this method has been widely employed to
explore the critical behavior of various models (e. g. \cite{zheng1998,zheng2001,
albano2001a, silva2002a, silva2002b, arashiro2003, silva2004a, grandi2004,
fernandes2005, hadjiagapiou2005, fernandes2006a, fernandes2006b,
fernandes2006c, prudnikov2010, silva2013a, silva2013b, silva2014,
chiocchetta2016, fernandes2017, silva2012, arashiro2007, luo2006,
silva2004b, silva2015, basu2017, loscar2016,loscar2017, albano2001b,
fernandes2016, albano2011}), highlighting its importance and effectiveness
in time-dependent MC simulations.

In models with absorbing states, a general scaling relation for the density
of active sites is given by \cite{hinrichsen2000,Hankel2010,marro1999}
\begin{equation}
\left\langle \rho (t)\right\rangle \sim t^{-\beta /\nu _{\parallel
}}f((y-y_{c})t^{1/\nu _{\parallel }},t^{d/z}L^{-d},\rho _{0}t^{\beta /\nu
_{\parallel }+\theta }),  \label{eq:fss}
\end{equation}%
where $\left\langle \cdots \right\rangle $ denotes the average over
different system evolutions, $\rho _{0}=\rho (0)$ is the density of active
sites at the start of the simulation, $d$ is the spatial dimension, $L$ is
the linear size of a square lattice, and $t$ represents the time evolution.
The indices $z=\nu _{\parallel }/\nu _{\perp }$ and $\theta =\frac{d}{z}-\frac{2\beta }{\nu _{\parallel }}$ are dynamic critical exponents, while $\beta $, $\nu _{\parallel }$, and $\nu _{\perp }$ are static critical
exponents. The control parameter $y$ corresponds to the CO-adsorption rate,
and $y_{c}$ is the critical CO-adsorption rate. Thus, $y-y_{c}$ represents
the distance from the critical point, which governs the algebraic behavior
of two independent correlation lengths: the spatial one, $\xi_{\perp }\sim
(y-y_{c})^{-\nu _{\perp }}$, and the temporal one, $\xi _{\parallel }\sim
(y-y_{c})^{-\nu _{\parallel }}$.

At criticality ($y=y_{c}$), it is possible to estimate different critical
exponents through the equation above when different conditions are
considered at the very beginning of the simulations ($t=0$). Thus, in the
context of time-dependent MC simulations, by starting from $\rho =\rho_{0}$
is expected the following crossover of power laws: 
\begin{equation}
\left\langle \rho (t)\right\rangle \sim \left\{ 
\begin{array}{ll}
\rho _{0}t^{\theta } & \text{if }t_{mic}<t<\rho _{0}^{-\frac{\nu _{\parallel
}}{d\ \nu _{\perp }-\beta }} \\ 
&  \\ 
t^{-\delta } & \text{if }\rho _{0}^{-\frac{\nu _{\parallel }}{d\ \nu _{\perp
}-\beta }}\leq t\text{,}%
\end{array}%
\right.  \label{Eq:Crossover}
\end{equation}%
with 
\begin{equation}
\theta =\frac{d}{z}-2\frac{\beta }{\nu _{\parallel }}\text{ and }\delta =%
\frac{\beta }{\nu _{\parallel }}\text{,}  \label{eq:exponents}
\end{equation}
Here, $t_{\min }$ represents a microscopic time scale that must be excluded
when performing Monte Carlo simulations. It is important to make a distinction here: in the case of the contact process, for instance, the minimum initial density for the system to evolve is $\rho _{0}=1/L^{2}$.

However, in the case of the ZGB model, even when all sites are initially
occupied by CO molecules, evolution will still occur in a system that
includes desorption. By using the density of vacant sites as the order
parameter, one can expect that when all sites on the lattice are initially vacant
(i. e., $\rho _{0}=\rho (0)=1$), the density of vacant sites will exhibit a
decay behavior $\left\langle \rho (t)\right\rangle \sim t^{-\delta }$, as
predicted by Eq. \ref{Eq:Crossover}.

On the other hand, if the sites are initially filled with CO molecules
(i.e., $\rho _{0}=0$), the order parameter is expected to follow $\left\langle \rho (t)\right\rangle \sim t^{\theta }$, which is also
predicted by the crossover behavior described in Eq. \ref{Eq:Crossover}. In
this case, $\delta $ and $\theta $ are expected, by hypothesis, to take the exact values
given by Eq. \ref{eq:exponents}.

In this study, we focus on characterizing an unprecedented critical line
identified in the ZGB model with CO desorption. Our analysis is restricted
to the power laws described by the crossover in Eq. \ref{Eq:Crossover}. To
localize the critical parameters, we employ a straightforward statistical
approach that proves highly efficient in monitoring temporal power laws
within the context of time-dependent Monte Carlo (MC) simulations: the
coefficient of determination (COD) \cite{trivedi2002}.

The concept itself recognizes that when performing a linear fit of
experimental data, there is an inherent source of variation, along with an
additional source of unexplained variation. The total variation is the sum
of both. The quality of a linear fit is then quantified as the ratio of
explained variation to total variation, yielding a value between 0 and 1.
This value is the COD, which is denoted as $r$ here. A higher $r$ indicates
a better linear fit. The method employed in this work is based on the
approach proposed in nonequilibrium Statistical Mechanics by da Silva,
Drugowich, and Martinez \cite{silva2012}, which used the coefficient of
determination in the linearization of the time evolution of the averaged
magnetizations to localize the critical parameters of the generalized Ising
model. However, this method is general and extends beyond this specific
application and has proven valuable in various other models \cite{fernandes2016,silva2013a, silva2013b,silva2014,silva2015,fernandes2017}.

We begin by considering the averages of the log-densities, $\left\langle \ln
\rho (t|y,k)\right\rangle $, across different runs. For each pair $(y,k)$,
we calculate $r$, the coefficient of determination, as:
\begin{equation}
r(y,k)=\frac{\sum\limits_{t=N_{\min }}^{N_{MC}}(\overline{\ln \left\langle
\rho (t|y,k)\right\rangle }-a-b\ln t)^{2}}{\sum\limits_{t=N_{\min
}}^{N_{MC}}(\overline{\ln \left\langle \rho (t|y,k)\right\rangle }-\ln
\left\langle \rho \right\rangle (t))^{2}}.  \label{eq:coef_det}
\end{equation}
Here, the overline denotes an additional average:
\begin{equation}
\overline{\ln \left\langle \rho (t|y,k)\right\rangle }=\frac{1}{N_{MC}}%
\sum\nolimits_{t=N_{\min }}^{N_{MC}}\ln \left\langle \rho \right\rangle (t).
\end{equation}

The numbers $N_{MC}$ and $N_{\min }$ represent, respectively, the total
number of MC steps used to obtain the results and the number of initial
steps discarded at the beginning of each simulation. Finally, the optimal
parameters are those obtained through marginalization:
\begin{eqnarray}
y_{opt} &=&\arg \max_{y_{\min }\leq y\leq y_{\max }}\{r(y|k)\} \\
&&  \nonumber \\
k_{opt} &=&\arg \max_{k_{\min }\leq k\leq k_{\max }}\{r(k|y)\}  \nonumber
\end{eqnarray}%
that is, by fixing $k$, we vary $y$ to identify the value that maximizes $r$, while by fixing $y$, we adjust $k$ to maximize $r$. The quantities $a$ and 
$b$ represent the intercept and slope, respectively, of the linear function.
When starting with $\rho _{0}=1$, $b$ corresponds to $\delta $ at the
critical value, while when $\rho _{0}=0$, it corresponds to $\theta $. The intervals $[y_{\min},y_{\max }]$ and $[k_{\min },k_{\max }]$ are appropriately chosen for the refinement.

As shown in Eq. (\ref{eq:coef_det}), $r$ represents the ratio of the
explained variation to the total variation. Thus, when the system is close
to the critical point ($y_{c},k_{c}$), $r\approx 1$, indicating that $\rho
(t)$ follows a power-law behavior, which, in turn, approaches a linear
function on a log-log scale. In contrast, when the system is away from the
critical point, $\rho (t)$ does not exhibit a power-law behavior as
described in Eq. \ref{Eq:Crossover}, and $r\approx 0$.

\section{Results}

\label{sec:results}

In this study, we explore the nonequilibrium critical behavior that arises
in the ZGB model when CO molecules are spontaneously desorbed from the
catalytic surface at a rate $k$. We demonstrate that the inclusion of
desorption reveals a hidden weak universality in the ZGB model, which we
uncover through time-dependent Monte Carlo simulations.

This further investigation has proved to be important in view of what we had
shown in Figure 1 of Ref. \cite{fernandes2018} which presents two plots of
the coefficient of determination $r$, Eq. (\ref{eq:coef_det}), as function
of $y$ and $k$. As can be seen in Figure 1 (b) of that paper, that little tail possesses two yellow lines (meaning $r \approx 1$) which eventually meet each other
around $y \approx 0.56$ and $k \approx 0.08$. Here, we are interested in
studying the bottom line which seems to arise around the first-order phase
transition of the standard ZGB model ($y \approx 0.525$ and $k=0$) whereas
we expect this transition not to hold for $k > 0$.

In our simulations, we consider square lattices with linear size $L=80$,
5,000 independent samples and run short-time MC simulations for the first $N_{MC}=300$ steps. In addition, the first $N_{\min }=50$ steps ($t_{mic}$ in
Eq. \ref{Eq:Crossover}) were discarded to avoid microscopic time scale
effects.

\begin{figure}[tbh]
\begin{center}
\includegraphics[width=1.0\columnwidth]{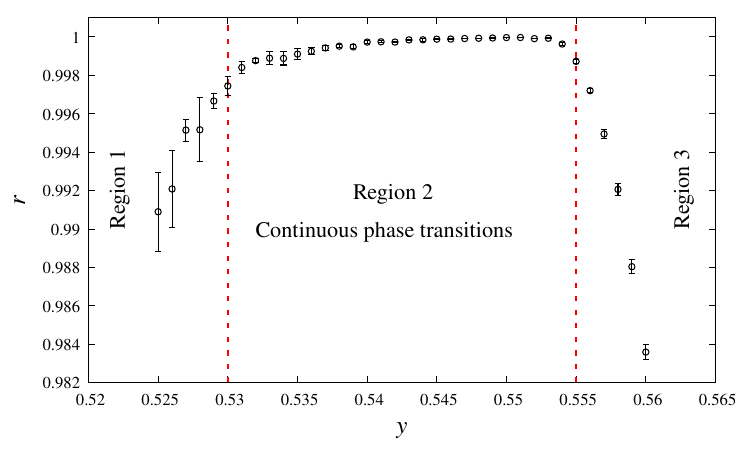}
\end{center}
\caption{Coefficient of determination $r$ as function of $y$ for $0.525\leq
y\leq 0.560$ and $\Delta y=10^{-3}$.}
\label{fig:r_x_y2}
\end{figure}

In order to achieve our goal, we first obtain the coefficient of
determination $r$ (Eq. (\ref{eq:coef_det})) around the yellow line described
above, for $y$ extending from 0.525 to 0.560 and $k$ from 0 to 0.08. In Fig. \ref{fig:r_x_y2} we present the behavior of $r$ as function of $y$ with $\Delta y=0.001$. 

\begin{figure}[tbh]
\begin{center}
\includegraphics[width=1.0\columnwidth]{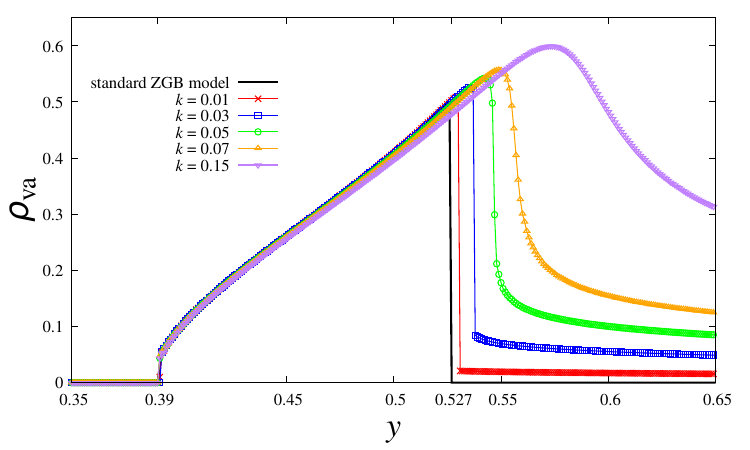} %
\includegraphics[width=1.0\columnwidth]{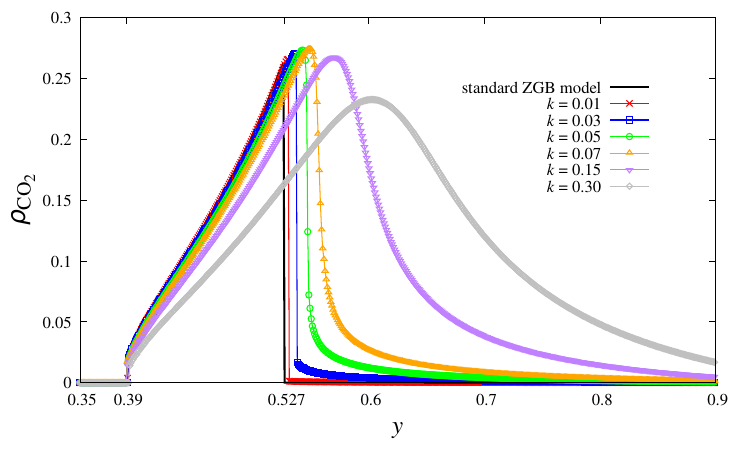}
\end{center}
\caption{Steady state Monte Carlo simulations for the ZGB model with
desorption of carbon dioxide ($\text{CO}_{2}$) molecules. Top: Density of
vacant sites, $\protect\rho _{\text{va}}$, as function of $y$ for different
values of $k$. Bottom: Density of carbon dioxide, $\protect\rho _{\text{CO}%
_{2}}$, as function of $y$ for different values of $k$.}
\label{fig:equilibrium}
\end{figure}

This result provided valuable insight into the behavior along this line,
prompting us to divide it into three distinct regions. In regions 1 and 3,
the points exhibit low coefficients of determination ($r \leq 0.997$),
indicating that these are likely not critical points. In region 1, the
points are situated near the first-order phase transition of the standard
ZGB model ($k \approx 0$ and $y \approx 0.525$), suggesting a strong
influence of the discontinuous transition on the system's behavior.
Conversely, in region 3, where $k > 0.065$, the existence of absorbing
states is effectively ruled out. This observation is further supported by
the steady-state simulation results shown in Fig. \ref{fig:equilibrium},
which we conducted to further support our results.

Figure \ref{fig:equilibrium} illustrates that the desorption of CO molecules
(at a rate $k$) from the catalytic surface does not affect the continuous
phase transition point of the standard ZGB model ($k = 0$ and $y \approx
0.389$), even at high desorption rates. However, as noted earlier, the
first-order phase transition is significantly influenced by increasing $k$.
For very small values of $k$ ($k \lesssim 0.03$), the abrupt changes in
density values indicate that the discontinuous phase transition
characteristic of the standard model continues to exert a strong influence,
potentially leading to a first-order-like or weak first-order phase
transition. This influence likely explains the low coefficients of
determination observed in region 1 of Fig. \ref{fig:r_x_y2}.

\begin{figure}[htb]
\begin{center}
\includegraphics[width=1.0\columnwidth]{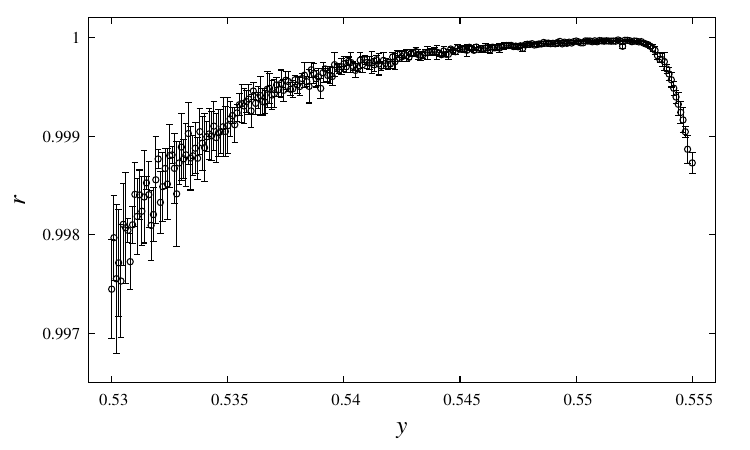}
\end{center}
\caption{Coefficient of determination $r$ as function of $y$ for $0.530 \leq
y \leq 0.555$ and $\Delta y=10^{-4}$.}
\label{fig:r_x_yfinal}
\end{figure}

On the other hand, the bottom panel of Fig. \ref{fig:equilibrium} provides
insight into region 3 of Fig. \ref{fig:r_x_y2}, as it reveals the
sustainable production of $\text{CO}_2$ molecules for $k \gtrsim 0.05$. In
this scenario, the active phase persists, and the model exhibits only a
continuous phase transition, separating the absorbing phase with O atoms
from the active phase. Consequently, the remainder of this paper will focus
exclusively on region 2 of Fig. \ref{fig:r_x_y2}, identified as the critical
region of the model. This region is characterized by $r \simeq 0.998$,
corresponding to $0.530 \leq y \leq 0.555$, as shown in Fig. \ref{fig:r_x_yfinal}.

This figure shows that the coefficient of determination increases from the
beginning as it moves away from the first-order phase transition point
approaching 1 for $y \simeq 0.553$ and suddenly drops, characterizing the
end of the critical region.

\begin{figure}[htb]
\begin{center}
\includegraphics[width=1.0\columnwidth]{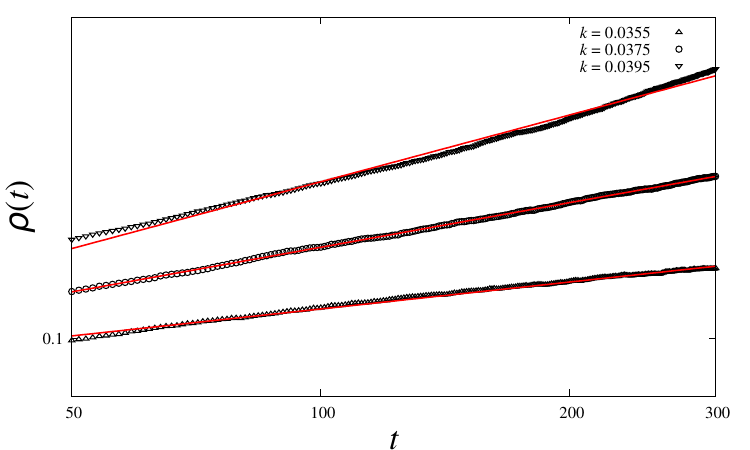}
\end{center}
\caption{Critical behavior of the density of vacant sites, $\protect\rho(t)$
as function of $t$ in $\log \times \log$ scale for $y_c=0.540$ and three
values of $k$: $k=0.0355$, 0.0375, and 0.0395.}
\label{fig:powerlaw_y0540}
\end{figure}

As an example of the use of the coefficient of determination, we present
Figure \ref{fig:powerlaw_y0540} which shows the linear behavior of the
density of vacant sites, $\rho(t)$, as function of $t$ (see the first power law of Eq. (\ref{Eq:Crossover})) in $\log \times \log$ scale for $y=0.540$ and three values of $k $: $k=0.0355$, 0.0375, and 0.0395.
This figure corroborates how much these curves deviate from the linear
behavior when the system is out of criticality. In other words, the closer
to criticality, the better the linear behavior of $\rho(t)$ and the higher the coefficient of determination. To this particular case, the best value
of the critical desorption rate for $y=y_c=0.540$ was $k=k_c=0.03754(17)$,
obtained when $r=0.99973(8)$.

\begin{figure}[htb]
\begin{center}
\includegraphics[width=1.0\columnwidth]{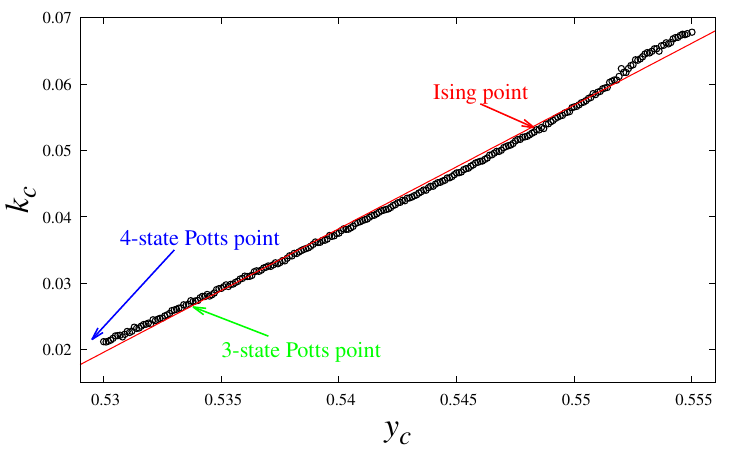}
\end{center}
\par
\caption{$k_c$ as function of $y_c$ for the critical region of the ZGB model with desorption of CO molecules. For clarity, we highlight key points related to the Ising model, as well as the 3-state and 4-state Potts models. The red line represents a linear fit showing the relationship between \( k_c \) and \( y_c \), which appears to hold for values of \( y_c \) below approximately 0.553.}
\label{fig:kc_x_yc}
\end{figure}

By gathering all the analysis and information presented above, we are now
able to show our estimates of the critical points along the critical region
of the model, $0.530 \leq y_c \leq 0.555$, as shown in Fig. \ref{fig:kc_x_yc}. 
As can be seen, the critical line obtained from the estimated critical points behaves approximately the same as a straight line, deviating of it only at the ends.

\begin{figure}[tbh]
\begin{center}
\includegraphics[width=1.0\columnwidth]{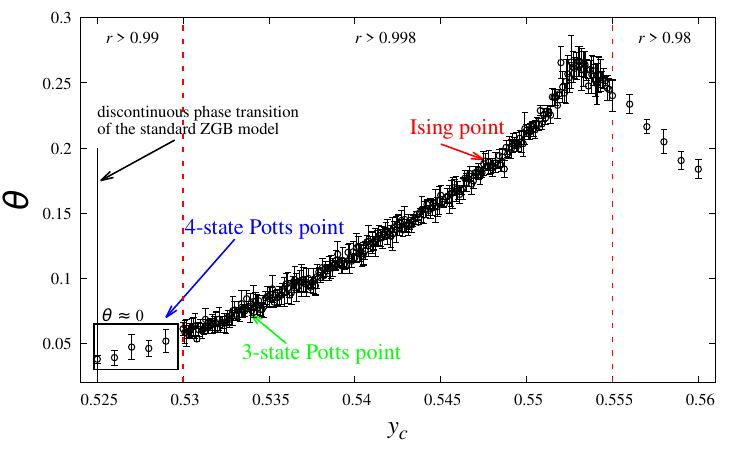}
\end{center}
\caption{The dynamic critical exponent \( \theta \) as a function of \( y_c \) along the critical line of the model.}
\label{fig:theta_x_yc}
\end{figure}

With these critical points in hand, we were able to obtain the critical
exponents of the model and from them, to observe if the universality class
of the standard ZGB model is maintained or not. Here, we decided to estimate
only the dynamic critical exponent $\theta $. Figure \ref{fig:theta_x_yc}
shows our estimates for several points along the critical line. 

As observed, the exponent $\theta$ varies along the critical line, indicating a weak universality class similar to the $Z_{5}$ and Ashkin-Teller models \cite{silva2014,fernandes2016}. The values range from $\theta \approx 0$ to $\theta \approx 0.27$, covering the exponents of the Ising model and the three- and four-state Potts models on two-dimensional square lattices.

It is important to clarify some details about the plot. All points in the region $0.53<y_{c}<0.555$ exhibit a good coefficient of determination, with $r > 0.99$. However, the central region displays an even better fit, with $r > 0.998$, corresponding to the data presented in Fig. \ref{fig:r_x_yfinal}. The region with smaller values of $\theta$ includes the first-order transition point of the standard ZGB model. Notably, in the vicinity of this point, $r$ is slightly lower than in the central region. Two important reasons are conjectured to explain this behavior:

\begin{enumerate}
\item A kind of "scar" from the first-order point of the original model reduces 
the coefficient of determination in this region;

\item The conjecture proposed in \cite{Okano1997} suggests that at the critical point of the four-state Potts model, $\theta$ is expected to be negative or very close to zero. This aligns with our simulation results at points approaching the discontinuity of the standard ZGB model. At the critical point of the standard four-state Potts model, the presence of a marginal operator generates significant noise in the region, making it difficult to identify power laws that accurately describe $\theta$, as noted in \cite{silva2004a}.
\end{enumerate}

Here, one could even conjecture that, similar to the two-dimensional symmetric Ashkin-Teller model, the starting point of this critical line may correspond to the four-state Potts point, given the significant similarities in the weak universality observed in both models.

%TCIMACRO{\TeXButton{B}{\begin{table}[tbp] \centering}}%
%BeginExpansion
\begin{table*}[tbp] \centering%
%EndExpansion
\begin{tabular}{ccccc}
\hline\hline
\textbf{Work} & \textbf{Model} & $y_{c}$ & $k_{c}$ & $\theta $ \\ 
\hline\hline
\multicolumn{1}{l}{Current work} & \multicolumn{1}{l}{%
\begin{tabular}{l}
Ising point via ZGB model%
\end{tabular}%
} & \multicolumn{1}{l}{%
\begin{tabular}{l}
$0.5476$%
\end{tabular}%
} & \multicolumn{1}{l}{%
\begin{tabular}{l}
$0.05166(22)$%
\end{tabular}%
} & \multicolumn{1}{l}{%
\begin{tabular}{l}
$0.190(7)$%
\end{tabular}%
} \\ 
\multicolumn{1}{l}{\cite{fernandes2018}} & \multicolumn{1}{l}{%
\begin{tabular}{l}
Ising point via ZGB model%
\end{tabular}%
} & \multicolumn{1}{l}{%
\begin{tabular}{l}
$0.554$%
\end{tabular}%
} & \multicolumn{1}{l}{%
\begin{tabular}{l}
$0.064$%
\end{tabular}%
} & \multicolumn{1}{l}{%
\begin{tabular}{l}
$0.198(1)$%
\end{tabular}%
} \\ 
\multicolumn{1}{l}{\cite{tome1993}} & \multicolumn{1}{l}{%
\begin{tabular}{l}
Ising point via ZGB model%
\end{tabular}%
} & \multicolumn{1}{l}{%
\begin{tabular}{l}
$0.54212(10)$%
\end{tabular}%
} & \multicolumn{1}{l}{%
\begin{tabular}{l}
$0.04060(5)$%
\end{tabular}%
} & $-$ \\ 
\multicolumn{1}{l}{\cite{chan2015}} & \multicolumn{1}{l}{%
\begin{tabular}{l}
Ising point via ZGB model%
\end{tabular}%
} & \multicolumn{1}{l}{%
\begin{tabular}{l}
$0.54052(9)$%
\end{tabular}%
} & \multicolumn{1}{l}{%
\begin{tabular}{l}
$0.0371(2)$%
\end{tabular}%
} & $-$ \\ 
\multicolumn{1}{l}{\cite{zheng1998,Okano1997}} & \multicolumn{1}{l}{%
\begin{tabular}{l}
Standard Ising model
\end{tabular}%
} & $-$ & $-$ & \multicolumn{1}{l}{%
\begin{tabular}{l}
$0.191(1)$%
\end{tabular}%
} \\ 
\multicolumn{1}{l}{\cite{TomeOliveira1998}} & \multicolumn{1}{l}{$%
\begin{tabular}{l}
Glauber Model \\ 
Majority vote model \\ 
Extreme Model%
\end{tabular}%
$} & $-$ & $-$ & \multicolumn{1}{l}{%
\begin{tabular}{l}
$0.191(2)$ \\ 
$0.190(5)$ \\ 
$0.188(8)$%
\end{tabular}%
} \\ 
\multicolumn{1}{l}{Current work} & \multicolumn{1}{l}{%
\begin{tabular}{l}
Three-state Potts point via ZGB model
\end{tabular}%
} & \multicolumn{1}{l}{%
\begin{tabular}{l}
$0.5353$%
\end{tabular}%
} & \multicolumn{1}{l}{%
\begin{tabular}{l}
$0.02950(23)$%
\end{tabular}%
} & \multicolumn{1}{l}{%
\begin{tabular}{l}
$0.082(4)$%
\end{tabular}%
} \\ 
\multicolumn{1}{l}{\cite{zheng1998}} & \multicolumn{1}{l}{%
\begin{tabular}{l}
Standard three-state Potts model
\end{tabular}%
} & $-$ & $-$ & \multicolumn{1}{l}{%
\begin{tabular}{l}
$0.075(3)$%
\end{tabular}%
} \\ 
\multicolumn{1}{l}{Current work} & \multicolumn{1}{l}{%
\begin{tabular}{l}
Four-state Potts point via ZGB model$^{\ast }$
\end{tabular}%
} & \multicolumn{1}{l}{%
\begin{tabular}{l}
$0.5250$%
\end{tabular}%
} & \multicolumn{1}{l}{%
\begin{tabular}{l}
$0.01318(30)$%
\end{tabular}%
} & \multicolumn{1}{l}{%
\begin{tabular}{l}
$0.038(3)$%
\end{tabular}%
} \\ 
\multicolumn{1}{l}{\cite{silva2004a,fernandes2006b}} & \multicolumn{1}{l}{%
\begin{tabular}{l}
Standard four-state Potts model
\end{tabular}%
} & $-$ & $-$ & \multicolumn{1}{l}{%
\begin{tabular}{l}
$-0.0429(11)$%
\end{tabular}%
} \\ \hline\hline
\end{tabular}%
\caption{Comparisons of our findings for $k_{c}$, $y_{c}$, and $\theta$, with those found in literature. The asterisk (*) in the four-state Potts model indicates that the exponent is estimated using \( y_c = 0.525 \), which corresponds to the discontinuous phase transition point of the standard ZGB model.}\label{tab:exponents}%
%TCIMACRO{\TeXButton{E}{\end{table}}}%
%BeginExpansion
\end{table*}%
%EndExpansion

Here, we conjecture an important phenomenon: when desorption is introduced ($k > 0$), the discontinuous transition is eliminated, but its "scar" persists, leading to the emergence of the four-state Potts point. This is not coincidental, since the Potts model with five or more states exhibits only first-order transitions, further supporting this interpretation. However, we emphasize that the specific mechanisms underlying this behavior require further exploration through alternative approaches and investigations.

In Table \ref{tab:exponents}, we present our estimates of the dynamic critical exponent $\theta$ for selected critical points, alongside results from the literature, allowing for direct comparisons. By analyzing the similarities in the exponent $\theta$, we can determine which $(y_{c},k_{c})$ corresponds to each point.

For instance, for the Ising model, our estimate of $\theta$ is $0.190(7)$, obtained 
with $y_{c}=0.5476$, which corresponds to $k_{c}=0.05166(22)$. This result aligns 
closely with previous estimates obtained by Tomé and Dickman \cite{tome1993} ($y_{c}=0.54212(10)$, $k_{c}=0.04060(5)$), Chan and Rickvold \cite{chan2015} ($y_{c}=0.54052(9)$, $k_{c}=0.0371(2)$), and in our previous work \cite{fernandes2018} ($y_c=0.554$, $k_c=0.064$).

Moreover, the exponent $\theta = 0.038(3)$ (near zero, as predicted in \cite{Okano1997} for the four-state Potts model) was obtained using the first-order transition point of the standard ZGB model. Specifically, $y_{c}=0.5250$ was used as input, resulting in $k_{c}=0.01318(30)$.

\begin{figure}[tbh]
\begin{center}
\includegraphics[width=1.0\columnwidth]{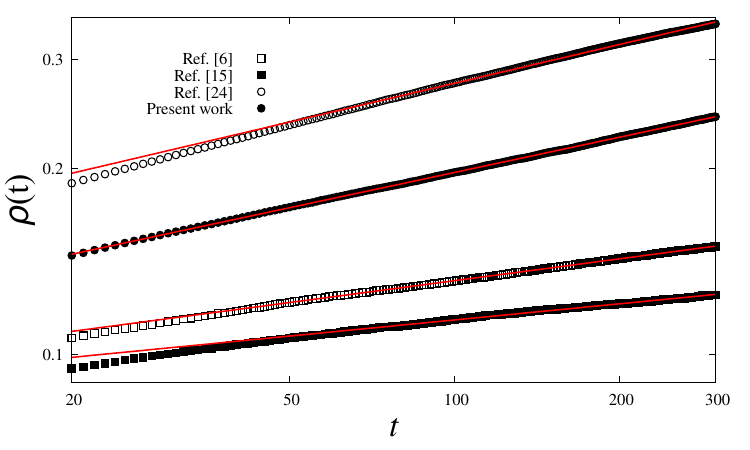}
\end{center}
\caption{Time evolution of the density of vacant sites for various Ising-like models, as reported in the literature}
\label{Fig:Different_Ising_like_points}
\end{figure}

Two final points warrant our attention. First, the Ising-like point has been observed by other authors and merits a visual comparison with our estimate. Figure \ref{Fig:Different_Ising_like_points} illustrates the different power laws obtained for the density of vacant sites of various studies.

It is worth noting that, aside from a method previously developed by the authors of this paper in another publication, the estimates found in the literature rely on distinct methodologies. Nonetheless, our results reveal a more robust power-law behavior starting from $t_{\text{mic}} = 10$ MC steps, particularly for our estimate of $y_{c} = 0.5476$ and $k_{c} = 0.05166(22)$. Our value of $\theta = 0.190(7)$ aligns closely with the best estimates for the Ising model point across different approaches.

While modern techniques make it easier to refine power-law behaviors by considering small deviations in $(y_{c},k_{c})$, it is crucial to recognize the foundational contributions of Rickvold and collaborators, as well as Tome and Dickman. Their pioneering work identified the Ising-like point through mean-field and steady-state Monte Carlo simulations. This groundwork was instrumental in guiding us toward uncovering the unprecedented critical line of weak universality, which is further analyzed in this study.

%Last but not least, it is important to highlight that our choice of lattice size is entirely appropriate. We conducted extensive simulations using our optimization method, which is faster than traditional steady-state simulations. Given the large number of points studied, selecting an optimal lattice size was essential; we determined $L = 80$ to be ideal. Figure \ref{Fig:finite_size_scaling} presents our estimates for the Ising-like point across different lattice sizes.

%\begin{figure}[tbh]
%\begin{center}
%\includegraphics[width=1.0\columnwidth]{fig_finite_size_scaling.png}
%\end{center}
%\caption{Time evolution of the density of vacant sites for the Ising-like
%point for our estimate, considering different size lattices. }
%\label{Fig:finite_size_scaling}
%\end{figure}

%The results clearly demonstrate that $L = 80$ is sufficient to mitigate issues related to finite-size scaling, as the power laws become consistent and uniform for lattices of this size and larger. The inset plot highlights the region for $t > 100$, showcasing the deviations observed for smaller lattice sizes ($L = 20$ and $L = 40$).

\section{Conclusions}

\label{sec:conclusions}

In this paper, we have performed time-dependent Monte Carlo simulations in the ZGB model with spontaneous desorption of CO molecules in order to show the existence of a critical line emerging from the discontinuous phase-transition point of the original model. Our findings support that this starting point may belong to the four-state Potts universality class. In addition, we have observed that this critical line passes through other very important critical points, such as the three-state Potts point and Ising point, confirming that this version of the model fits into a very particular class of Statistical Mechanic models which possess weak universality. We believe that these findings will shed new light on previously unexplored aspects of criticality in the ZGB model. 

%In this paper, we demonstrate the existence of a previously unobserved line of weak universality in the Ziff-Gulari-Barshad (ZGB) model with desorption. Our findings are derived from time-dependent Monte Carlo simulations optimized for the coefficient of determination and further corroborated by steady-state Monte Carlo simulations.

%We show that the disappearance of the first-order transition point due to desorption not only gives rise to an Ising-like critical point, as predicted in the literature, but also reveals a continuum of critical points. These points span from the Ising-like point through the 3-state Potts-like point to the 4-state Potts point, which we interpret as a "scar" of the vanished first-order transition.

%Our results are characterized by the universality of the dynamic exponent $\theta$, extensively analyzed in the context of short-time dynamics. We believe these findings will shed new light on previously unexplored aspects of criticality in the ZGB model. 

\section*{Acknowledgments}

RDS and HAF thanks Conselho Nacional de Desenvolvimento Científico e Tecnológico of Brazil (CNPq) for supporting this work under grants number 304575/2022-4 and 405508/2021-2, respectively.

\end{document}